\documentclass[aps,prl,showpacs,twocolumn,superscriptaddress,letterpaper]{revtex4-1}

\usepackage{graphicx}
\usepackage{dcolumn}
\usepackage{bm}
\usepackage{amsmath, amssymb}%

\begin{document}


\title{Short Range Correlations and the EMC Effect}

%
%
%
%
%
%
%
%
%
%

\newcommand*{\ODU}{Old Dominion University, Norfolk, Virginia 23529}
\newcommand*{\ODUindex}{1}
\affiliation{\ODU}
\newcommand*{\TAU}{Tel Aviv University, Tel Aviv 69978, Israel}
\newcommand*{\TAUindex}{2}
\affiliation{\TAU}
\newcommand*{\JLAB}{Thomas Jefferson National Accelerator Facility, Newport News, Virginia 23606}
\newcommand*{\JLABindex}{3}
\affiliation{\JLAB}
 

  
\author{L.B.~Weinstein}
\email[Contact Author \ ]{weinstein@odu.edu}
     \affiliation{\ODU}
\author{E.~Piasetzky}
\affiliation{\TAU}
\author{D.W. Higinbotham}
\affiliation{\JLAB}
\author{J. Gomez}
\affiliation{\JLAB}
\author{O. Hen}
\affiliation{\TAU}
\author{R. Shneor}
\affiliation{\TAU}

\date{\today}

\begin{abstract}
  This paper shows quantitatively that the magnitude of the EMC effect
  measured in electron deep inelastic scattering (DIS) at intermediate
  $x_B$, $0.35\le x_B\le 0.7$, is linearly related to the Short Range
  Correlation (SRC) scale factor obtained from electron inclusive
  scattering at $x_B\ge 1$.  The observed
  phenomenological relationship is used to extract the ratio of the
  deuteron to the free $pn$ pair cross sections and $F_2^n/F_2^p$, the ratio of the free neutron to
  free proton structure functions.  We speculate that the observed
  correlation is because both the EMC effect and SRC are dominated
  by the high virtuality (high momentum) nucleons in the nucleus.
\end{abstract}


\maketitle


Inclusive electron scattering, $A(e,e')$, is a valuable
tool for studying nuclei.  By selecting specific kinematic conditions,
especially the four-momentum and energy transfers, $Q^2$ and $\omega$,
one can focus on different aspects of the nucleus.  Elastic scattering
has been used to measure the nuclear charge distribution. Deep
inelastic scattering at $Q^2 > 2$ GeV$^2$, and $0.35\le x_B \le 0.7$
($x_B=Q^2/2m\omega$, where $m$ is the nucleon mass) is sensitive
to the nuclear quark distributions. Inelastic scattering at $Q^2 >
1.4$ GeV$^2$ and $x_B > 1.5$ is sensitive to nucleon-nucleon short
range correlations (SRC) in the nucleus.  This paper will explore the
relationship between deep inelastic and large-$x_B$ inelastic
scattering.

The per-nucleon electron deep inelastic scattering (DIS) cross
sections of nuclei with $A\ge 3$ are smaller than those of deuterium at
$Q^2 \ge 2$ GeV$^2$, and moderate $x_B$, $0.35 \le x_B \le 0.7$. This
effect, known as the EMC effect, has been measured for a wide range of
nuclei 
\cite{Aubert83,Ashman88,Gomez94,Arneodo88,Arneodo90,Allasia90,Seely09}. There
is no generally accepted explanation of the EMC effect. In general,
proposed explanations need to include both nuclear structure effects
(momentum distributions and binding energy) and modification of the
bound nucleon structure due to the nuclear medium. Comprehensive
reviews of the EMC effect can be found in
\cite{Geesaman95,Norton03,Sargsian:2002wc,Smith:2002ci} and references
therein. Recent high-precision data on light nuclei \cite{Seely09}
suggest that it is a local density effect and not a bulk property of
the nuclear medium.

The per-nucleon electron inelastic scattering cross sections of nuclei
with $A\ge 3$ are greater than those of deuterium for $Q^2>1.4$
GeV$^2$ and large $x_B$, $1.5 \le x_B \le 2$. The cross section ratio
for two different nuclei ({\it e.g.}, carbon and helium) shows a
plateau when plotted as a function of $x_B$ ({\it i.e.,} it is
independent of $x_B$).  This was first observed at SLAC \cite{Frankfurt93}
and subsequently at Jefferson Laboratory \cite{Egiyan03,Egiyan06}.
The plateau indicates that the nucleon momentum distributions of
different nuclei for high momentum, $p\ge p_{thresh} = 0.275$ GeV/c, are similar in
shape and differ only in magnitude.  The ratio (in the plateau region)
of the per-nucleon inclusive $(e,e')$ cross sections for two nuclei is
then the ratio of the probabilities to find high momentum nucleons in
those two nuclei \cite{Frankfurt81,Frankfurt88}.

These high-momentum nucleons were shown recently in hadronic
\cite{Tang03,Piasetzky06} and leptonic \cite{Shneor07,Subedi08}
two-nucleon knockout experiments to be almost entirely due to central
and tensor nucleon-nucleon Short Range Correlations (SRC)
\cite{sargsian05,schiavilla07,alvioli08,baghdasaryan10}. SRC occur between pairs
of nucleons with high relative momentum and low center of mass
momentum, where low and high are relative to the Fermi momentum in heavy
nuclei.  Thus, we will call the ratio of cross sections in the plateau
region the ``SRC scale factor''.

This paper will show quantitatively that the magnitude of the EMC
effect in nucleus $A$ is linearly related to the SRC scale factor of
that nucleus relative to deuterium.  This idea was suggested by
Higinbotham {\it et al.}  \cite{Higinbotham10}.

We characterize the strength of the EMC effect for nucleus $A$
following Ref.~\cite{Seely09} as the slope of the ratio of the
per-nucleon deep inelastic electron scattering cross sections of
nucleus $A$ relative to deuterium, $dR_{EMC}/dx$, in the region
$0.35\le x_B\le0.7$.  This slope is proportional to the value of the
cross section ratio at $x\approx 0.5$, but is unaffected by overall
normalization uncertainties that merely raise or lower all of the data
points together.  For $^3$He, $^4$He, $^9$Be and $^{12}$C we use the
published slopes from \cite{Seely09} measured at $3\le Q^2 \le 6$
GeV$^2$. We also fit the ratios,
measured in Ref. \cite{Gomez94}, as a function of $x_B$ for $0.36\le
x_B \le 0.68$. The results are averages over all measured $Q^2$ ({\it
  i.e.,} $Q^2=2,5$ and 10 GeV$^2$ for $x_B<0.6$ and $Q^2=5$ and 10
GeV$^2$ for larger $x_B$).  The results from the two measurements for
$^4$He and $^{12}$C are consistent and we use the weighted
average of the two.  See Table \ref{tab:EMC}.  The uncertainties are
not meant to take into account possible effects of the anti-shadowing
region at $x_B\approx 0.15$ and the Fermi motion region at $x_B>0.75$
extending into the region of interest.

\begin{table}[htb]
\begin{center} 
\begin{tabular}{| l | c | c | c |} 
\hline 
& $dR_{EMC}/dx$ & $dR_{EMC}/dx$ & $dR_{EMC}/dx$ \\ 
Nucleus & (Ref. \cite{Seely09}) & (Ref. \cite{Gomez94}) & (combined) \\ \hline 
Deuteron & & & 0 \\ \hline
$^3$He & $-0.070\pm0.029$ & & $-0.070\pm0.029$ \\ \hline 
$^4$He & $-0.199\pm0.029$ & $-0.191\pm0.061$ & $-0.197\pm0.026$ \\ \hline 
$^9$Be & $-0.271\pm0.029$ & $-0.207\pm0.037$ & $-0.243\pm0.023$ \\ \hline 
$^{12}$C & $-0.280\pm0.029$ & $-0.318\pm0.040$ & $-0.292\pm0.023$  \\\hline 
$^{27}$Al & & $-0.325\pm0.034$ & $-0.325\pm0.034$ \\ \hline
$^{40}$Ca & &  $-0.350\pm0.047$ & $-0.350\pm0.047$ \\ \hline
$^{56}$Fe & & $-0.388\pm0.032$ & $-0.388\pm0.032$ \\ \hline 
$^{108}$Ag & & $-0.496\pm0.051$  & $-0.496\pm0.051$ \\ \hline
$^{197}$Au & & $-0.409\pm0.039$  & $-0.409\pm0.039$ \\ \hline
\end{tabular} 
\caption{The measured EMC slopes $dR_{EMC}/dx$ for $0.35\le x_B\le 0.7$. \label{tab:EMC}}
\end{center}
\end{table}

The SRC scale factors determined from the isospin-corrected
per-nucleon ratio of the inclusive
$(e,e')$ cross sections on nucleus $A$ and $^3$He, $a_2(A/{}^3{\rm
  He}) = (3/A)(\sigma_A(Q^2,x_B)/\sigma_{^3{\rm He}}(Q^2,x_B)$
are listed in Table \ref{tab:SRC} using data from \cite{Egiyan06}.
We used the ratio of deuterium to $^3$He determined in
Ref.~\cite{Egiyan06} primarily from the calculated ratio of their
momentum distributions above the scaling threshold
($p_{thresh}=0.275\pm0.025$ GeV/c). We
combined the statistical and systematic uncertainties in quadrature to
give the total uncertainties shown in the table.  The SRC scale
factors for nucleus $A$ relative to deuterium, $a_2(A/d)$,
are calculated from the second column. 

The value of the SRC scale factors was shown to be $Q^2$ independent
for $1.5 \le Q^2 \le 2.5$ GeV$^2$ \cite{Egiyan03} and more recently
for $1.5 \le Q^2 \le 5$ GeV$^2$ \cite{Fomin07}.  Similarly, the EMC
effect was shown to be $Q^2$ independent for SLAC, BCDMS and NMC data
for $2 \le Q^2 \le 40$ GeV$^2$ \cite{Gomez94}.  This $Q^2$-independence
allows us to compare these quantities in their different measured
ranges.

\begin{table}[htb]
\begin{center} 
\begin{tabular}{| l | l | l |l|} 
\hline 
 & Measured & Measured & Predicted \\
Nucleus & $a_2(A/{}^3{\rm He})$ & $a_2(A/d)$ & $a_2(A/d)$ \\ \hline 
Deuteron & $0.508\pm0.025$ & 1 & \\ \hline
$^3$He & 1 & $1.97\pm0.10$ & \\ \hline 
$^4$He & $1.93\pm0.14$ & $3.80\pm0.34$ & \\ \hline 
$^{12}$C & $2.41\pm0.17$ & $4.75\pm0.41$ & \\ \hline 
$^{56}$Fe & $2.83\pm0.18$ & $5.58\pm0.45$ & \\ \hline \hline
$^9$Be & & & $4.08\pm0.60$ \\ \hline
$^{27}$Al & & & $5.13\pm0.55$ \\ \hline
$^{40}$Ca & & & $5.44\pm0.70$ \\ \hline
$^{108}$Ag & & & $7.29\pm0.83$ \\ \hline
$^{197}$Au & & & $6.19\pm0.65$ \\ \hline
\end{tabular} 
\caption{The SRC scale factors for nucleus $A$ with respect to $^3$He and to
  deuterium. The third column is calculated from the second.  The resulting uncertainties
are slightly overestimated since the uncertainty in the $d$/$^3$He ratio
of about 5\% is added to all of the other ratios. The predicted
values (fourth column) are calculated from the values in Table
\ref{tab:EMC} and Eq.~\ref{eq:slope}.  \label{tab:SRC}}
\end{center}
\end{table}

\begin{figure}[htb]
\begin{center}
\includegraphics[scale=0.40]{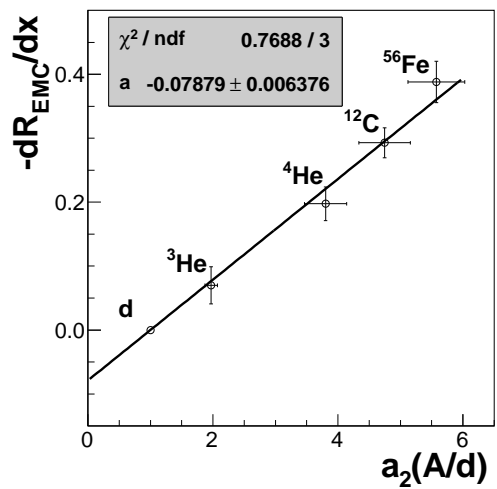}
\caption{ \label{fig:SRCEMC} The EMC slopes versus the SRC scale
  factors.  The uncertainties include both statistical and systematic
  errors added in quadrature.  The fit parameter is the intercept of
  the line and also the negative of the slope of the line.}
\end{center}
\end{figure}

Fig.~\ref{fig:SRCEMC} shows the EMC slopes versus the SRC scale
factors. The two values are strongly linearly correlated,
\begin{equation}
-dR_{\rm EMC}/dx = (a_2(A/d) - 1)\times (0.079\pm0.006)\quad.
\label{eq:slope}
\end{equation}
This implies that both
stem from the same underlying nuclear physics, such as high local
density or large nucleon virtuality ($v = P^2 - m^2$ where $P$ is the
four momentum).

This striking correlation means that we can predict the SRC scale
factors for a wide range of nuclei from Be to Au using the linear
relationship from Eq.~\ref{eq:slope} and the measured EMC slopes (see
Table~\ref{tab:SRC}).  Note that $^9$Be is a particularly interesting
nucleus because of its cluster structure and because its EMC slope is
much larger than that expected from a simple dependence on average
nuclear density \cite{Seely09}.  The EMC slopes and hence the
predicted SRC scale factors may saturate for heavy nuclei but better
data are needed to establish the exact $A$ dependence.

This correlation between the EMC slopes and the SRC scale
factors also allows us to extract significant information about the
deuteron itself.  Due to the lack of a free neutron target, the EMC
measurements used the deuteron as an approximation to a free proton
and neutron system and measured the ratio of inclusive DIS on nuclei
to that of the deuteron. This seems like a reasonable approximation
since the deuteron is loosely bound ($\approx 2$ MeV) and the average
distance between the nucleons is large ($\approx 2$ fm). But the
deuteron is not a free system; the pion tensor force binds the two
nucleons even if weakly.

To quantify the effects of the binding of nucleons in deuterium, 
we define the In-Medium Correction (IMC) effect as the ratio of the
DIS cross section per nucleon bound in a nucleus relative to the free
(unbound) $pn$ pair cross section (as opposed to the EMC effect which
uses the ratio to deuterium).

The deuteron IMC effect can be extracted from the data in
Fig.~\ref{fig:SRCEMC}.  If the IMC effect and the SRC scale factor
both stem from the same cause, then the IMC effect and the SRC scale
factor will both vanish at the same point. The value $a_2(A/d)=0$ is
the limit of free nucleons with no SRC. Extrapolating the best fit
line in Fig.~\ref{fig:SRCEMC} to $a_2(A/d)=0$ gives an intercept of
$dR_{EMC}/dx = -0.079\pm0.006$.  The difference between this value and
the deuteron EMC slope of 0 is the deuteron IMC slope:
\begin{equation}
\Bigg\vert\frac{dR_{IMC}(d)}{dx}\Bigg\vert = 0.079\pm0.006 \quad .
\end{equation}
This slope is the same size as the EMC slope measured for the ratio of
$^3$He to deuterium \cite{Seely09}. It is slightly smaller than the
deuterium IMC slope of $\approx0.10$ derived in \cite{Gomez94}
assuming that the EMC effect is proportional to the average nuclear
density and the slope of 0.098 deduced by Frankfurt and Strikman
based on the relative virtuality of nucleons in iron and deuterium
\cite{Frankfurt88} and the iron EMC slope \cite{Gomez94}.

The IMC effect for nucleus $A$ is then just the difference
between the measured EMC effect and the value
$dR_{EMC}/dx=-0.079\pm0.006$.  Thus
\begin{equation}
\Bigg\vert\frac{dR_{IMC}(A)}{dx}\Bigg\vert = \Bigg\vert\frac{dR_{EMC}(A)}{dx}\Bigg\vert_{meas} + 0.079\pm0.006
\quad .\end{equation}
This is true when the slopes are small compared to one.

The free neutron
cross section  can be obtained from the measured deuteron and proton cross sections
using the observed phenomenological relationship presented in
Fig. \ref{fig:SRCEMC} to determine the nuclear corrections.  Since the EMC effect is linear for $0.3\le x_B \le
0.7$, we have
\begin{equation}
\frac{\sigma_d}{\sigma_p + \sigma_n} = 1 - a(x_B - b) \quad \hbox{for} \quad 0.3 \le x_B \le 0.7,
\label{eq:sigd}
\end{equation}
where $\sigma_d$ and $\sigma_p$ are the measured DIS cross sections
for the deuteron and free proton, $\sigma_n$ is the free neutron DIS cross
section that we want to extract, $a = \vert
dR_{IMC}(d)/dx\vert=0.079\pm0.006$ and $b=0.31\pm0.04$ is the
average value of $x_B$ where the EMC ratio is unity ({\it i.e.,} where
the per-nucleon cross sections are equal $\sigma_A(x_B)/A =
\sigma_d(x_B)/2$) as determined in Refs.~\cite{Gomez94,Seely09} and taking into
account the quoted normalization uncertainties.  

Our results imply that $\sigma_d/(\sigma_p+\sigma_n)$ decreases
linearly from 1 to 0.97 over the range $0.3 \le x_B \le 0.7$.  (More
precisely, that it decreases by $0.031\pm0.004$ where the uncertainty
is due to the fit uncertainties in Eq. 3.) This depletion (see Eq.{} \ref{eq:sigd}) is similar in size to the depletion calculated by Melnitchouk using
the weak binding approximation smearing function with target mass
corrections and an off-shell correction \cite{Melnitchouk10}.
However, the distribution in $x_B$ is very different.  Melnitchouk's
calculated ratio reaches its minimum of 0.97 at $x_B\approx0.5$ and
increases rapidly, crossing 1 at $x_B\approx 0.7$.

If the structure function $F_2$ is proportional to the DIS cross
section ({\it i.e.,} if the ratio of the longitudinal to
transverse cross sections is the same for $n,p$ and $d$ [see
discussion in \cite{Geesaman95}]),
then the free neutron structure function, $F_2^n(x_B,Q^2)$, can also
be deduced from the measured deuteron and proton structure functions:
\begin{equation}
F_2^n(x_B,Q^2) = \frac{2F_2^d(x_B,Q^2) -
  [1-a(x_B-b)]F_2^p(x_B,Q^2)}{[1-a(x_B-b)]}
\end{equation}
which leads to
\begin{equation}
\frac{F_2^n(x_B,Q^2)}{F_2^p(x_B,Q^2)} = \frac{2\frac{F_2^d(x_B,Q^2)}{F_2^p(x_B,Q^2)} -
  [1-a(x_B-b)]}{[1-a(x_B-b)]} \quad .
\end{equation}
This is only valid for $0.35\le x_B \le 0.7$.  

\begin{figure}[htb]
\begin{center}
\includegraphics[scale=0.3]{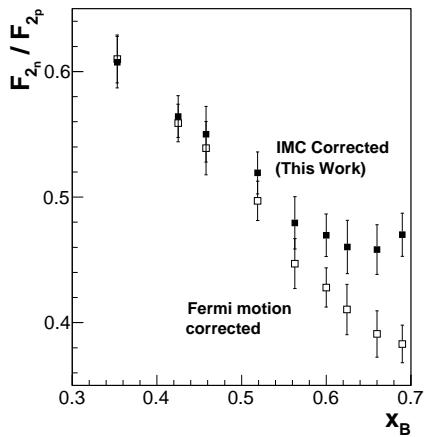}
\caption{ \label{fig:F2} The ratio of neutron to proton structure
  functions, $F_2^n(x_B,Q^2)/F_2^p(x_B,Q^2)$ as extracted from the
  measured deuteron and proton structure functions, $F_2^d$ and
  $F_2^p$. The filled symbols show $F_2^n/F_2^p$ extracted in this
  work from the deuteron In Medium Correction (IMC) ratio and the
  world data for $F_2^d/F_2^p$ at $Q^2 = 12$ GeV$^2$
  \cite{Arrington09}. The open symbols show $F_2^n/F_2^p$ extracted
  from the same data correcting only for nucleon motion in deuterium using a relativistic deuteron momentum density
  \cite{Arrington09}.}
\end{center}
\end{figure}

Fig.~\ref{fig:F2} shows the ratio of $F_2^n/F_2^p$ extracted in this
work using the IMC-based correction and the $Q^2=12$ GeV$^2$ ratio
$F_2^d/F_2^p$ from Ref.~\cite{Arrington09}.  Note that the
ratio $F_2^d/F_2^p$ is $Q^2$-independent from $6 \le Q^2 \le 20$
GeV$^2$ for $0.4\le x_B \le 0.7$ \cite{Arrington09}. 
The dominant uncertainty
in this extraction is the uncertainty in the measured $F_2^p/F_2^d$.
The IMC-based correction increases the extracted free neutron
structure function (relative to that extracted using the deuteron
momentum density \cite{Arrington09}) by an amount that increases with
$x_B$.  Thus, the IMC-based $F_2^n$ strongly favors model-based extractions of
$F_2^n$ that include nucleon modification in the deuteron \cite{Melnitchouk96}.  

The IMC-based $F_2^n$ appears to be constant or slightly increasing in
the range from $0.6 \le x_B \le 0.7$.  The $d/u$ ratio is simply related to the ratio of
$F_2^n/F_2^p$ in the deep inelastic limit, $x^2 \ll Q^2/4m^2$
\cite{Arrington09}, $d/u = (4F_2^n/F_2^p - 1)/(4-F_2^n/F_n^p)$.  While it is quite hazardous to
extrapolate from our limited $x_B$ range all the way to $x_B = 1$, these
results appear to disfavor models of the proton with $d/u$ ratios of
0 at $x_B=1$ (see \cite{Melnitchouk96} and references therein).  

By using the deuteron IMC slope, these results take into account both
the nuclear corrections as well as any possible changes to the
internal structure of the neutron in the deuteron.  Note that this
assumes either that the EMC and $F_2$ data are taken at the same $Q^2$
or that they are $Q^2$-independent for $6 \le Q^2 \le 12$ GeV$^2$.   The fact that
the measured EMC ratios for nuclei with $A\ge 3$ decrease linearly
with increasing $x_B$ for $0.35\le x_B \le 0.7$ indicates that Fermi
smearing is not significant in this range.

We now speculate as to the physical
reason for the EMC-SRC relation presented above. 
Assuming that the IMC/EMC effect is due to a difference in the
quark distributions in bound and free nucleons, these differences
could occur predominantly in either mean field nucleons or in nucleons
affected by SRC.


According to Ref.~\cite{Ciofi07}, the IMC/EMC effect is mainly associated
with nucleons at high virtuality.  These nucleons, like
the nucleons affected by SRC, have larger momenta and a denser
local environment than that of the other nucleons in the
nucleus. Therefore, they should exhibit the largest changes in their
internal structure.

The linear correlation between the strength of the EMC and the SRC in
nuclei, shown in Fig.~\ref{fig:SRCEMC},  indicates that possible
modifications of the quark distributions occur in nucleons affected by
SRC.  This also predicts a larger EMC effect in
higher density nuclear systems such as neutron stars. This correlation
may also help us to understand the difficult to quantify nucleon modification
(offshell effects) that must occur when two nucleons are close
together.

To summarize, we have found a striking linear correlation between the
EMC slope measured in deep inelastic electron scattering and the short
range correlations scale factor measured in inelastic scattering.  The
SRC are associated with large nucleon momenta and the EMC effect is
associated wth modified nucleon structure.  This correlation allows us
to extract the free neutron structure function model-independently and
to place constraints on large $x_B$ parton distribution functions.
Knowledge of these PDFs is important for searches for new
physics in collider experiments \cite{Kuhlmann00} and for neutrino
oscillation experiments.

We are grateful for many fruitful discussions with John Arrington,
Sebastian Kuhn, Mark Strikman, Franz Gross, Jerry Miller, and Wally Melnitchouk.  This
work was supported by the U.S. Department of Energy, the U.S. National
Science Foundation, the Israel Science Foundation, and the US-Israeli
Bi-National Science Foundation. Jefferson Science Associates operates
the Thomas Jefferson National Accelerator Facility under DOE contract
DE-AC05-06OR23177.


\begin{thebibliography}{32}
\expandafter\ifx\csname natexlab\endcsname\relax\def\natexlab#1{#1}\fi
\expandafter\ifx\csname bibnamefont\endcsname\relax
  \def\bibnamefont#1{#1}\fi
\expandafter\ifx\csname bibfnamefont\endcsname\relax
  \def\bibfnamefont#1{#1}\fi
\expandafter\ifx\csname citenamefont\endcsname\relax
  \def\citenamefont#1{#1}\fi
\expandafter\ifx\csname url\endcsname\relax
  \def\url#1{\texttt{#1}}\fi
\expandafter\ifx\csname urlprefix\endcsname\relax\def\urlprefix{URL }\fi
\providecommand{\bibinfo}[2]{#2}
\providecommand{\eprint}[2][]{\url{#2}}

\bibitem[{\citenamefont{Aubert et~al.}(1983)}]{Aubert83}
\bibinfo{author}{\bibfnamefont{J.}~\bibnamefont{Aubert}} \bibnamefont{et~al.},
  \bibinfo{journal}{Phys. Lett. B} \textbf{\bibinfo{volume}{123}},
  \bibinfo{pages}{275} (\bibinfo{year}{1983}).



\bibitem[{\citenamefont{Ashman et~al.}(1988)}]{Ashman88}
\bibinfo{author}{\bibfnamefont{J.}~\bibnamefont{Ashman}} \bibnamefont{et~al.},
  \bibinfo{journal}{Phys. Lett. B} \textbf{\bibinfo{volume}{202}},
  \bibinfo{pages}{603} (\bibinfo{year}{1988}).



\bibitem[{\citenamefont{Gomez et~al.}(1994)}]{Gomez94}
\bibinfo{author}{\bibfnamefont{J.}~\bibnamefont{Gomez}} \bibnamefont{et~al.},
  \bibinfo{journal}{Phys. Rev. D} \textbf{\bibinfo{volume}{49}},
  \bibinfo{pages}{4348} (\bibinfo{year}{1994}).



\bibitem[{\citenamefont{Arneodo et~al.}(1988)}]{Arneodo88}
\bibinfo{author}{\bibfnamefont{M.}~\bibnamefont{Arneodo}} \bibnamefont{et~al.},
  \bibinfo{journal}{Phys. Lett. B} \textbf{\bibinfo{volume}{211}},
  \bibinfo{pages}{493} (\bibinfo{year}{1988}).



\bibitem[{\citenamefont{Arneodo et~al.}(1990)}]{Arneodo90}
\bibinfo{author}{\bibfnamefont{M.}~\bibnamefont{Arneodo}} \bibnamefont{et~al.},
  \bibinfo{journal}{Nucl. Phys. B} \textbf{\bibinfo{volume}{333}},
  \bibinfo{pages}{1} (\bibinfo{year}{1990}).



\bibitem[{\citenamefont{Allasia et~al.}(1990)}]{Allasia90}
\bibinfo{author}{\bibfnamefont{D.}~\bibnamefont{Allasia}} \bibnamefont{et~al.},
  \bibinfo{journal}{Phys. Lett. B} \textbf{\bibinfo{volume}{249}},
  \bibinfo{pages}{366} (\bibinfo{year}{1990}).



\bibitem[{\citenamefont{Seely et~al.}(2009)}]{Seely09}
\bibinfo{author}{\bibfnamefont{J.}~\bibnamefont{Seely}} \bibnamefont{et~al.},
  \bibinfo{journal}{Phys. Rev. Lett.} \textbf{\bibinfo{volume}{103}},
  \bibinfo{pages}{202301} (\bibinfo{year}{2009}).



\bibitem[{\citenamefont{Geesaman et~al.}(1995)\citenamefont{Geesaman, Saito,
  and Thomas}}]{Geesaman95}
\bibinfo{author}{\bibfnamefont{D.}~\bibnamefont{Geesaman}},
  \bibinfo{author}{\bibfnamefont{K.}~\bibnamefont{Saito}}, \bibnamefont{and}
  \bibinfo{author}{\bibfnamefont{A.}~\bibnamefont{Thomas}},
  \bibinfo{journal}{Ann. Rev. Nucl. and Part. Sci.}
  \textbf{\bibinfo{volume}{45}}, \bibinfo{pages}{337} (\bibinfo{year}{1995}).



\bibitem[{\citenamefont{Norton}(2003)}]{Norton03}
\bibinfo{author}{\bibfnamefont{P.~R.} \bibnamefont{Norton}},
  \bibinfo{journal}{Rep. Prog. Phys.} \textbf{\bibinfo{volume}{66}},
  \bibinfo{pages}{1253} (\bibinfo{year}{2003}).


\bibitem{Sargsian:2002wc}
   M.~M.~Sargsian {\it et al.},
   J.\ Phys.\ G {\bf 29}, R1 (2003).


\bibitem{Smith:2002ci}
   J.~R.~Smith and G.~A.~Miller,
   Phys.\ Rev.\  C {\bf 65}, 055206 (2002).

\bibitem{Frankfurt93} L.L. Frankfurt and M.I. Strikman and D.B. Day
  and M. Sargsyan, Phys.\ Rev.\ C {\bf 48}, 2451 (1993).





\bibitem[{\citenamefont{Egiyan et~al.}(2003)}]{Egiyan03}
\bibinfo{author}{\bibfnamefont{K.}~\bibnamefont{Egiyan}} \bibnamefont{et~al.},
  \bibinfo{journal}{Phys. Rev. C} \textbf{\bibinfo{volume}{68}},
  \bibinfo{pages}{014313} (\bibinfo{year}{2003}). 



\bibitem[{\citenamefont{Egiyan et~al.}(2006)}]{Egiyan06}
\bibinfo{author}{\bibfnamefont{K.}~\bibnamefont{Egiyan}} \bibnamefont{et~al.},
  \bibinfo{journal}{Phys. Rev. Lett.} \textbf{\bibinfo{volume}{96}},
  \bibinfo{pages}{082501} (\bibinfo{year}{2006}). 



\bibitem[{\citenamefont{Frankfurt and Strikman}(1981)}]{Frankfurt81}
\bibinfo{author}{\bibfnamefont{L.~L.} \bibnamefont{Frankfurt}}
  \bibnamefont{and} \bibinfo{author}{\bibfnamefont{M.~I.}
  \bibnamefont{Strikman}}, \bibinfo{journal}{Phys. Rep.}
  \textbf{\bibinfo{volume}{76}}, \bibinfo{pages}{215} (\bibinfo{year}{1981}).



\bibitem[{\citenamefont{Frankfurt and Strikman}(1988)}]{Frankfurt88}
\bibinfo{author}{\bibfnamefont{L.}~\bibnamefont{Frankfurt}} \bibnamefont{and}
  \bibinfo{author}{\bibfnamefont{M.}~\bibnamefont{Strikman}},
  \bibinfo{journal}{Phys. Rep.} \textbf{\bibinfo{volume}{160}},
  \bibinfo{pages}{235 } (\bibinfo{year}{1988}).


\bibitem[{\citenamefont{Tang et~al.}(2003)}]{Tang03}
\bibinfo{author}{\bibfnamefont{A.}~\bibnamefont{Tang}} \bibnamefont{et~al.},
  \bibinfo{journal}{Phys. Rev. Lett.} \textbf{\bibinfo{volume}{90}},
  \bibinfo{pages}{042301} (\bibinfo{year}{2003}). 



\bibitem[{\citenamefont{Piasetzky et~al.}(2006)\citenamefont{Piasetzky,
  Sargsian, Frankfurt, Strikman, and Watson}}]{Piasetzky06}
\bibinfo{author}{\bibfnamefont{E.}~\bibnamefont{Piasetzky}},
  \bibinfo{author}{\bibfnamefont{M.}~\bibnamefont{Sargsian}},
  \bibinfo{author}{\bibfnamefont{L.}~\bibnamefont{Frankfurt}},
  \bibinfo{author}{\bibfnamefont{M.}~\bibnamefont{Strikman}}, \bibnamefont{and}
  \bibinfo{author}{\bibfnamefont{J.~W.} \bibnamefont{Watson}},
  \bibinfo{journal}{Phys. Rev. Lett.} \textbf{\bibinfo{volume}{97}},
  \bibinfo{pages}{162504} (\bibinfo{year}{2006}).



\bibitem[{\citenamefont{Shneor et~al.}(2007)}]{Shneor07}
\bibinfo{author}{\bibfnamefont{R.}~\bibnamefont{Shneor}} \bibnamefont{et~al.},
  \bibinfo{journal}{Phys. Rev. Lett.} \textbf{\bibinfo{volume}{99}},
  \bibinfo{eid}{072501} (\bibinfo{year}{2007}).



\bibitem[{\citenamefont{Subedi et~al.}(2008)}]{Subedi08}
\bibinfo{author}{\bibfnamefont{R.}~\bibnamefont{Subedi}} \bibnamefont{et~al.},
  \bibinfo{journal}{Science} \textbf{\bibinfo{volume}{320}},
  \bibinfo{pages}{1476} (\bibinfo{year}{2008}).



\bibitem[{\citenamefont{Sargsian et~al.}(2005)\citenamefont{Sargsian,
  Abrahamyan, Strikman, and Frankfurt}}]{sargsian05}
\bibinfo{author}{\bibfnamefont{M.~M.} \bibnamefont{Sargsian}},
  \bibinfo{author}{\bibfnamefont{T.~V.} \bibnamefont{Abrahamyan}},
  \bibinfo{author}{\bibfnamefont{M.~I.} \bibnamefont{Strikman}},
  \bibnamefont{and} \bibinfo{author}{\bibfnamefont{L.~L.}
  \bibnamefont{Frankfurt}}, \bibinfo{journal}{Phys. Rev.}
  \textbf{\bibinfo{volume}{C71}}, \bibinfo{pages}{044615}
  (\bibinfo{year}{2005}).



\bibitem[{\citenamefont{Schiavilla et~al.}(2007)\citenamefont{Schiavilla,
  Wiringa, Pieper, and Carlson}}]{schiavilla07}
\bibinfo{author}{\bibfnamefont{R.}~\bibnamefont{Schiavilla}},
  \bibinfo{author}{\bibfnamefont{R.~B.} \bibnamefont{Wiringa}},
  \bibinfo{author}{\bibfnamefont{S.~C.} \bibnamefont{Pieper}},
  \bibnamefont{and} \bibinfo{author}{\bibfnamefont{J.}~\bibnamefont{Carlson}},
  \bibinfo{journal}{Phys. Rev. Lett.} \textbf{\bibinfo{volume}{98}},
  \bibinfo{eid}{132501} (\bibinfo{year}{2007}).



\bibitem[{\citenamefont{Alvioli et~al.}(2008)\citenamefont{Alvioli, degli Atti,
  and Morita}}]{alvioli08}
\bibinfo{author}{\bibfnamefont{M.}~\bibnamefont{Alvioli}},
  \bibinfo{author}{\bibfnamefont{C.} \bibnamefont{Ciofi degli Atti}},
  \bibnamefont{and} \bibinfo{author}{\bibfnamefont{H.}~\bibnamefont{Morita}},
  \bibinfo{journal}{Phys. Rev. Lett.} \textbf{\bibinfo{volume}{100}},
  \bibinfo{eid}{162503} (\bibinfo{year}{2008}).


\bibitem[{\citenamefont{Baghdasaryan et~al.}(2010)}]{baghdasaryan10}
\bibinfo{author}{\bibfnamefont{H.}~\bibnamefont{Baghdasaryan}}
  \bibnamefont{et~al.}   \bibinfo{journal}{Phys. Rev. Lett.} \textbf{\bibinfo{volume}{105}},
  \bibinfo{eid}{222501}  (\bibinfo{year}{2010}).





\bibitem[{\citenamefont{Higinbotham et~al.}(2010)\citenamefont{Higinbotham,
  Gomez, and Piasetzky}}]{Higinbotham10}
\bibinfo{author}{\bibfnamefont{D.~W.} \bibnamefont{Higinbotham}},
  \bibinfo{author}{\bibfnamefont{J.}~\bibnamefont{Gomez}}, \bibnamefont{and}
  \bibinfo{author}{\bibfnamefont{E.}~\bibnamefont{Piasetzky}}
  (\bibinfo{year}{2010}), {\tt arXiv:1003.4497 [hep-ph]}.



\bibitem[{\citenamefont{Fomin}(2007)}]{Fomin07}
\bibinfo{author}{\bibfnamefont{N.}~\bibnamefont{Fomin}}, Ph.D. thesis,
  \bibinfo{school}{{University of Virginia}} (\bibinfo{year}{2007}),
  {\tt arXiv:0812.2144 [nucl-ex]}.




\bibitem[{\citenamefont{Melnitchouk}(2010)}]{Melnitchouk10}
\bibinfo{author}{\bibfnamefont{W.}~\bibnamefont{Melnitchouk}},
  \bibinfo{journal}{AIP Conf. Proc.} \textbf{\bibinfo{volume}{1261}},
  \bibinfo{pages}{85} (\bibinfo{year}{2010}), {\tt arXiv:1006.4134 [nucl-th]}.



\bibitem[{\citenamefont{Arrington et~al.}(2009)\citenamefont{Arrington,
  Coester, Holt, and Lee}}]{Arrington09}
\bibinfo{author}{\bibfnamefont{J.}~\bibnamefont{Arrington}},
  \bibinfo{author}{\bibfnamefont{F.}~\bibnamefont{Coester}},
  \bibinfo{author}{\bibfnamefont{R.}~\bibnamefont{Holt}}, \bibnamefont{and}
  \bibinfo{author}{\bibfnamefont{T.-S.~H.} \bibnamefont{Lee}},
  \bibinfo{journal}{J. Phys. G} \textbf{\bibinfo{volume}{36}},
  \bibinfo{pages}{025005} (\bibinfo{year}{2009}).






\bibitem[{\citenamefont{Melnitchouk and Thomas}(1996)}]{Melnitchouk96}
\bibinfo{author}{\bibfnamefont{W.}~\bibnamefont{Melnitchouk}} \bibnamefont{and}
  \bibinfo{author}{\bibfnamefont{A.~W.} \bibnamefont{Thomas}},
  \bibinfo{journal}{Phys. Lett. B} \textbf{\bibinfo{volume}{377}},
  \bibinfo{pages}{11} (\bibinfo{year}{1996}).









\bibitem[{\citenamefont{{C. Ciofi degli Atti, L.L. Frankfurt, L.P. Kaptari and
  M.I. Strikman}}(2007)}]{Ciofi07}
\bibinfo{author}{\bibnamefont{{C. Ciofi degli Atti, L.L. Frankfurt, L.P.
  Kaptari and M.I. Strikman}}}, \bibinfo{journal}{Phys. Rev. C}
  \textbf{\bibinfo{volume}{76}}, \bibinfo{pages}{055206}
  (\bibinfo{year}{2007}).
  




\bibitem[{\citenamefont{S. Kuhlmann et~al.}(2000)}]{Kuhlmann00}
\bibinfo{author}{\bibnamefont{S.~Kuhlmann}} \bibnamefont{et~al.},
 \bibinfo{journal}{Phys. Lett. B} \textbf{\bibinfo{volume}{476}},
 \bibinfo{pages}{291} (\bibinfo{year}{2000}).


\end{thebibliography}

\end{document}